\documentclass[aps, pra, reprint,superscriptaddress,nofootinbib]{revtex4-1}

\usepackage{graphicx}
\usepackage{dcolumn}
\usepackage{bm}
\usepackage{amsfonts}
\usepackage{amsmath}
\usepackage{amssymb}
\usepackage{color}

\usepackage[colorlinks=true,citecolor=blue]{hyperref}
\hypersetup{colorlinks=true,citecolor=blue,linkcolor=red,urlcolor=blue}

\renewcommand{\thefigure}{\arabic{figure}}
\def\be{\begin{equation}}
\def\ee{\end{equation}}

\def\kv{{\bf k}}
\def\qv{{\bf q}}

\def\vv{{\bf v}}
\def\jv{{\bf j}}

\def\zhat{{\hat z}}
\def\sigmav{{\bm \sigma}}
\def\tauv{{\bm \tau}}
\newcommand{\h}[1]{{\hat {#1}}}
\newcommand{\hdg}[1]{{\hat {#1}^\dagger}}

\begin{document}

\title{Many-body correction to the intrinsic anomalous and spin Hall conductivities}
\author{Moslem Mir}
\email{moslem.mir@uoz.ac.ir}
	\affiliation{Department of Physics, University of Zabol (UOZ), Zabol 98615-538, Iran}
\author{Saeed H. Abedinpour}
\email{abedinpour@iasbs.ac.ir}
	\affiliation{Department of Physics, Institute for Advanced Studies in Basic Sciences (IASBS), Zanjan 45137-66731, Iran}
	\affiliation{Research Center for Basic Sciences \& Modern Technologies (RBST), Institute for Advanced Studies in Basic Sciences (IASBS), Zanjan 45137-66731, Iran}

\date{\today}

%%%%%%%%%%%%%%%%%%%%%%%%
\begin{abstract}
Broken Galilean invariance in a spin-orbit coupled system can amplify many-body effects on its different responses. We study the anomalous Hall and spin Hall conductivities of a magnetic two-dimensional electron gas with Rashba spin-orbit coupling. We show that both of these conductivities in the intrinsic limit are fully specified in terms of the longitudinal and transverse spin-spin response functions. 
We include the effect of electron-electron interaction in the spin-spin linear response functions, going beyond the random-phase approximation. We do this by incorporating the local-field correction in the response functions, which takes into account the many-body exchange-correlation effects.
We observe a significant enhancement of the static anomalous Hall conductivity due to the electron-electron interaction. 
The many-body correction on the spin Hall effects is more non-trivial, and strong electron-electron interaction can even reverse the sign of the static spin Hall conductivity.
\end{abstract}
\maketitle

%%%%%%%%%%%%%%%%%%%
\emph{Introduction.}--
The anomalous Hall and spin Hall effects are two prominent examples in the realm of the Hall effects,  which have attracted immense theoretical and experimental attention. 
In the anomalous Hall effect (AHE), an in-plane electric field produces a transverse charge current in the absence of an external magnetic field~\cite{nagaosa_rmp2010}. The spin Hall effect (SHE) refers to the generation of a transverse spin-current in response to an in-plane electric field~\cite{Sinova2015}.
The experimental detection of anomalous Hall conductivity is straightforward, as it simply requires measurement of the transverse electric potential difference. The spin Hall conductivity, on the other hand, could be detected using different experiments~\cite{Sinova2015}, in particular, Kerr rotation microscopy~\cite{Hernandez2013}, electrical measurement~\cite{Garlid2010} and quantum interferences~\cite{Werake2011}.

For both anomalous Hall and spin Hall effects, spin-orbit coupling (SOC) plays an essential role~\cite{Hibino}, and several competing contributions make the complete theoretical understanding of these phenomena quite complicated and sometimes with a lot of controversy~\cite{raimondi_annalen2012}.
Different microscopic contributions to AHE and SHE are usually divided into intrinsic and extrinsic mechanisms~\cite{nagaosa_rmp2010,Hankiewicz2008}.
As the name suggests, the intrinsic contribution originates from the material's electronic structure. The extrinsic mechanism refers to the scattering of the charge carriers from impurities and includes side-jump and skew scatterings~\cite{Sinova2015}. 
In some simple analytically treatable models, an exact cancellation between intrinsic and extrinsic parts is predicted~\cite{Dimitrova2005}. However, note that many-body corrections are generaly not considered in such studies.
In a non-interacting two-dimensional electron gas with Rashba SOC, extrinsic and intrinsic contributions to the spin Hall conductivity are expected to entirely cancel each other~\cite{Sinova2015}.
In a magnetic Rashba system, Ado et al.~\cite{Ado_PRL2016} showed that incorporating the contribution of non-crossing diagrams in the skew scattering breaks the exact cancellation between the intrinsic and extrinsic contributions to the anomalous Hall conductivity.

In most conventional systems, such as the regular electron liquids~\cite{GV_book}, the Galilean invariance protects many response functions from the many-body corrections at long wavelengths.  
Since spin-orbit coupling breaks the Galilean invariance~\cite{Amit2011, Maslov2015, Maslov2016, Maslov2017, Mir},  we naturally expect to find traces of electron-electron interaction in both anomalous and spin Hall conductivities. 
While a vast body of theoretical studies has explored different aspects of AHE and SHE, the influence of electron-electron interaction on these conductivities is less explored~\cite{Hankiewicz_PRB2006, Amit2011}. 
The main challenge in treating the many-body effects in the Hall conductivities is that one needs to include the effects of many-body exchange-correlations going beyond the well-established and widely used random phase approximation (RPA), whereas doing so is not always straightforward. 

In this letter, we aim to include the effects of electron-electron interaction in the intrinsic dynamical anomalous and spin Hall conductivities. We consider a magnetic two-dimensional (2D) electron gas with a linear Rashba spin-orbit coupling. This is a simple, analytically treatable platform where both inversion and time-reversal symmetries are broken; therefore, we can anticipate finite anomalous Hall and spin Hall responses. 
Furthermore, we are interested in the high frequency or clean limit; hence we can safely discard extrinsic contributions to these conductivities ~\cite{Hankiewicz2008}. Such ballistic processes are expected to be the dominant transport mechanisms in the nanoscale spintronic devices~\cite{Werake2011}.   

In the following, we begin with introducing our model and defining the main quantities. Then using the equation of motion method for the current-current linear response functions, we show that the intrinsic conductivities are fully specified in terms of the longitudinal and transverse spin-spin linear response functions. 
This result is generic and reproduces all the known results in the noninteracting regime, as well as in the interacting non-magnetic two-dimensional electron gas with Rashba SOC. 
We include many-body effects in the spin-spin response functions utilizing a Hubbard-like many-body local-field correction~\cite{Mir}, as a first step to go beyond the RPA. In this way, the effect of electron-electron interaction on the anomalous and spin Hall conductivities is also revealed.

\emph{Model and general definitions.}--   
We consider an interacting two-dimensional electron gas with Rashba spin-orbit coupling and a perpendicular exchange field. The Hamiltonian of this system reads
\be\label{H_Full}
H=H_0+H_{\rm int},
\ee
where
\be\label{hamilH_{0}}
H_0=\sum_{\kv,\sigma,\sigma'}\varepsilon_{\sigma\sigma'}(k)\hdg{c}_{\kv,\sigma}\h{c}_{\kv,\sigma'},
\ee
is the single-particle part of the Hamiltonian. Here, $\hat{c}^{\dagger}_{\kv,\sigma}$ and $\hat{c}_{\kv,\sigma}$ create and annihilate an electron with wave vector $\kv$ and spin $\sigma$, respectively. 
The single-particle energy matrix in the spin-space reads ($\hbar=1$)
\be
\varepsilon(k)=\frac{ k^2}{2 m}\tau^0
+\alpha_{\rm R}\left( \zhat \times  \tauv \right)\cdot \kv+\Delta \tau^z,
\ee
where $m$, $\alpha_{\rm R}$, and $\Delta$ are the electron mass, Rashba spin-orbit coupling strength, and Zeeman energy splitting due to the exchange field, respectively, $\zhat$ is a unit vector in the direction perpendicular to the 2D plane, $\tau^{\alpha}$ refers to the identity ($\alpha=0$) and three Pauli matrices ($\alpha=x,y,z$), and the Pauli-matrix vector is $\tauv\equiv(\tau^x,\tau^y,\tau^z)$.
The Zeeman term can either come from an external exchange field by means of a Ferromagnetic sub-lattice or originate from the spontaneous symmetry breaking of the Rashba electron gas due to the electron-electron interaction~\cite{Liu_PRB2017,Liu_PRB2020}.
The second term on the right-hand-side of Eq.~\eqref{H_Full}, $H_{\rm int}$, is the electron-electron interaction term 
\be\label{H_int}
H_{\rm int}=\frac{1}{2S}\sum_{\qv\neq 0} v(q)\sum_{\kv,\kv'}\sum_{\sigma,\sigma'}\hdg{c}_{\kv-\qv,\sigma}  \hdg{c}_{\kv'+\qv,\sigma'}\h{c}_{\kv',\sigma'} \h{c}_{\kv,\sigma}.
\ee 
Here, $S$ is the sample area, and $v(q)=2\pi e^2/q$ is the Fourier transform of the bare Coulomb interaction in two dimensions.
$H_{0}$ is easily diagonalized to give the single-particle energy bands
\be\label{e_k}
\varepsilon_{k,\mu}=\frac{ k^2}{2m}+\mu \sqrt{\Delta^2+\alpha_{\rm R}^2 k^2},
\ee
and the eigenvectors
\be\label{vectors}
\psi_{\kv,\mu} =\left(\begin{array}{cc} 
\sin(\phi_{k,\mu}/2)\\ 
-i\mu  \cos(\phi_{k,\mu}/2) e^{i\theta_{k}}
\end{array}\right) , 
\ee
where $\mu=\pm 1$ is the band index, $\theta_{k}$ is the angle between $\kv$ and $x$-axis,
and defining $\varphi_{k}\equiv 2 \arctan[{\alpha_{\rm R} k}/({\sqrt{\Delta^2+\alpha_{\rm R}^2 k^2}-\Delta})]$,
we have $\phi_{k,+}=\phi_k$ and $\phi_{k,-}=\pi-\phi_{k}$.

We will use current and spin-current operators to define the anomalous and spin Hall conductivities; so, they are introduced below.
The electron velocity is given by $\vv=\kv/m+\alpha_{\rm R}(\zhat \times \tauv)$~\cite{Marinescu2007}. Therefore the total (i.e., $q=0$) \emph{paramagnetic} current density operator reads
\be\label{J-para}
\hat{\jv}
=\frac{\hat {\bf P}_{\rm CM}}{m}+\alpha_{\rm R}(\zhat \times \hat{\sigmav}_{\rm tot}).
\ee
Here, the center-of-mass (CM) momentum is defined as 
\be
\hat{\bf P}_{\rm CM}\equiv\sum_{\kv,\sigma} \kv \hdg{c}_{\kv,\sigma}\h{c}_{\kv,\sigma},
\ee
and the total spin vector is
\be
 \hat{\sigmav}_{\rm tot}\equiv\sum_{\kv,\sigma,\sigma'}\hat{c}^{\dagger}_{\kv,\sigma}\tauv_{\sigma\sigma'}\hat{c}_{\kv,\sigma'}.
\ee
Note that, as we are not considering vector potentials here, we will not make any distinction between the paramagnetic and physical current operators~\cite{GV_book, Amit2011}.
Similarly, the z-component of the total spin-current operator~\cite{Marinescu2007} $\jv^z= \{\vv,\tau^z\}/4= \kv/(2m) \tau^z$, is written as
\be
\hat{\jv}^{z}=\sum_{\kv,\sigma,\sigma'}\frac{ \kv}{2m}\hat{c}^{\dagger}_{\kv,\sigma}\tau^{z}_{\sigma\sigma'}\hat{c}_{\kv,\sigma'}.
\ee
Furthermore, with the help of the Heisenberg equation of motion for the total spin operator $i\partial_t \hat {\bf A}=[\hat{\bf A},H]$, where $H$ is the full Hamiltonian of the system given by Eq.~\eqref{H_Full}, we can write
\be\label{j-spin}
\hat{\jv}^z=-\frac{1}{4m \alpha_{\rm R}}\partial_t  \hat{\sigmav}_{\rm tot}+\frac{\Delta}{2m \alpha_{\rm R}} (\zhat\times  \hat{\sigmav}_{\rm tot}).
\ee 

%%%%%%%%%%%%%%%%%%%%%%%%%%%%%%%%%
\emph{Dynamical anomalous and spin Hall conductivities.}--
The anomalous Hall and spin Hall conductivities, respectively, describe transverse charge and spin-polarized currents in response to a homogeneous (i.e., $q=0$) electric field. Considering a two-dimensional system in the $xy$-plane, if the electric field $E$ is taken in the $x$-direction, then the $y$-components of the particle current $j_y$ and $z$-polarized spin-current $j_y^z$, respectively read
\be
\begin{split}
j_{y}(\omega)&=\sigma^{\rm AH}_{yx}(\omega)E_{x}(\omega),\\
j^{z}_{y}(\omega)&=\sigma^{\rm SH}_{yx}(\omega)E_{x}(\omega).
\end{split}
\ee
Here, $\sigma^{\rm AH}_{yx}(\omega)$ is the dynamical anomalous Hall conductivity
\be\label{A-H-C}
\sigma^{\rm AH}_{yx}(\omega)=\frac{ie^2}{\omega}\chi_{j_yj_x}(\omega),
\ee
and $\sigma^{\rm SH}_{yx}(\omega)$ is the dynamical spin Hall conductivity
\be\label{S-H-C}
\sigma^{\rm SH}_{yx}(\omega)=\frac{ie}{\omega}\chi_{j^z_yj_x}(\omega),
\ee
where 
\be
\begin{split}
\chi_{AB}(\omega)&=\frac{1}{S}\langle\langle \hat{A};\hat{B} \rangle\rangle_\omega\\
&\equiv -\frac{i}{S}\int_0^\infty \mathrm{d}t \langle [\hat{A}(t),\hat{B}(0)]\rangle e^{i(\omega+i \eta) t},
\end{split}
\ee
is the retarded linear response function~\cite{GV_book}, with $\eta$ an infinitesimal positive value. 
For the anomalous Hall conductivity, we make use of Eq.~\eqref{J-para} for the current operator to find
\be\label{jyjx}
\begin{split}
\langle\langle \hat{j}_y; \hat{j}_x \rangle\rangle_\omega &=
\langle\langle \frac{\hat{P}_{{\rm CM},y}}{m}+\alpha_{\rm R}\sigma^x_{\rm tot}; \frac{\hat{P}_{{\rm CM},x}}{m}-\alpha_{\rm R}\sigma^y_{\rm tot} \rangle\rangle_\omega\\
&=-\alpha_{\rm R}^2\langle\langle \sigma^x_{\rm tot}; \sigma^y_{\rm tot} \rangle\rangle_\omega,
\end{split}
\ee
where we have used the fact that total momentum is conserved in the clean or high-frequency limit $\omega \tau \rightarrow \infty$ (here, $\tau$ is the electron-impurity scattering lifetime)~\cite{Brune2010,Hernandez2013,Amit2011}.

Now, inserting Eq.~\eqref{jyjx} in Eq.~\eqref{A-H-C} we find
\be\label{A-H-C-IN-B}
\sigma^{\rm AH}_{yx}(\omega)=\frac{ie^2\alpha^2_{\rm R}}{\omega}\chi_{yx}(\omega),
\ee
where $\chi_{\alpha\beta}(\omega)\equiv\langle\langle  \sigma^\alpha_{\rm tot} ;\sigma^\beta_{\rm tot} \rangle\rangle_\omega/S$, is the spin-spin linear response function, and we have made use of the reciprocity relation $\chi_{yx}(\omega)=-\chi_{xy}(\omega)$.

Similarly, for the spin Hall conductivity, from Eqs.~\eqref{J-para} and~\eqref{j-spin}, we can write
\be\label{jzyjx}
\begin{split}
\langle\langle \hat{j}^z_y; \hat{j}_x \rangle\rangle_\omega &=
\langle\langle \frac{-\partial_t \sigma^y_{\rm tot}+2\Delta\sigma^x_{\rm tot}}{4 m \alpha_{\rm R}}; \frac{\hat{P}_{{\rm CM},x}}{m}-\alpha_{\rm R}\sigma^y_{\rm tot} \rangle\rangle_\omega\\
&=\frac{1}{4m}\langle\langle \partial_t\sigma^y_{\rm tot}; \sigma^y_{\rm tot} \rangle\rangle_\omega
-\frac{\Delta}{2m}\langle\langle \sigma^x_{\rm tot}; \sigma^y_{\rm tot} \rangle\rangle_\omega,
\end{split}
\ee
where again, the conservation of the total momentum is employed. 
The second term on the right-hand-side of the above equation is the spin-spin response function, while for the first term, we can make use of the identity
$\omega\langle \langle \h{A};\h{B}\rangle \rangle_{\omega}=\langle[\h{A},\h{B}]\rangle+i\langle \langle \partial_t \hat{A};\h{B}\rangle \rangle_{\omega}$~\cite{GV_book, Amit2011}, to write
\be\label{partial_sy}
\langle\langle \partial_t\sigma^y_{\rm tot}; \sigma^y_{\rm tot} \rangle\rangle_\omega=
-i\omega
\langle \langle \sigma^y_{\rm tot}; \sigma^y_{\rm tot} \rangle \rangle_\omega.
\ee
Now, upon the substitution of Eq.~\eqref{partial_sy} in Eq.~\eqref{jzyjx}, and then the obtained result in Eq.~\eqref{S-H-C}, the spin Hall conductivity reads 
\be\label{S-H-C-IN-B}
\sigma^{\rm SH}_{yx}(\omega)=\frac{e}{4m }\chi_{yy}(\omega)+\frac{ie\Delta}{2m \omega} \chi_{yx}(\omega).
\ee
We would like to note that equations \eqref{A-H-C-IN-B} and \eqref{S-H-C-IN-B} for the intrinsic dynamical anomalous Hall and spin Hall conductivities, in terms of the spin-spin linear response functions, are our main formal results.  
Both expressions are valid as long as the clean system limit behavior is of interest.
Also notice that the term proportional to $i \chi_{yx}(\omega)/\omega$ in Eqs.~\eqref{A-H-C-IN-B} and \eqref{S-H-C-IN-B}, in the $\omega\to 0$ limit, is related to the transverse spin diffusion constant. 
In the absence of an exchange field (i.e., $\Delta=0$), only the first term on the right-hand-side of Eq.~\eqref{S-H-C-IN-B}, 
\be\label{S-H-C-IN}
\left.\sigma^{{\rm SH}}_{yx}(\omega)\right|_{\Delta=0}=\frac{e}{4m }\chi_{yy}(\omega),
\ee
survives, which is the noted expression for the dynamical spin Hall conductivity of non-magnetic 2D Rashba system in terms of the in-plane susceptibility~\cite{Dimitrova2005,Shekhter,Amit2011,Mir}.

Before discussing the effects of electron-electron interaction on the spin responses and Hall conductivities, we will briefly review the anomalous and spin Hall effects in a noninteracting system.

%%%%%%%%%%%%%%%%%%%%%%%%%%%%%%%%%%%
\emph{Noninteracting anomalous and spin Hall conductivities.}--
In the noninteracting limit, we simply need to use the noninteracting spin-spin response functions on the right-hand sides of Eqs.~\eqref{A-H-C-IN-B} and \eqref{S-H-C-IN-B}. 
For the noninteracting anomalous Hall conductivity, we find~\cite{Supplemental}
\be\label{A-H-C-Non}
\sigma^{{\rm AH},0}_{yx}(\omega)=\frac{ie^2\alpha_{\rm R}^2}{\omega}\chi^{0}_{yx}(\omega)
=\frac{e^2\Delta}{4\pi \omega}L(\omega),
\ee
where
\be\label{eq:lw}
L(\omega)=\log\left[\frac{(\omega-\omega_{-}+i0^{+})(\omega+\omega_{+}+i0^{+})}{(\omega-\omega_{+}+i0^{+})(\omega+\omega_{-}+i0^{+})}\right],
\ee
with $\omega_{\pm}=2\sqrt{\Delta^2+2\varepsilon_{\rm F}m\alpha_{\rm R}^2+m^2\alpha_{\rm R}^4} \mp 2m\alpha_{\rm R}^2$. Here, $\varepsilon_{\rm F}$ is the Fermi energy of the system, which we take to be greater than $\Delta$ throughout this paper, so the system is always in the two-band regime.
In the static limit, using the low-frequency expansion of $L(\omega)$~\cite{Supplemental}, we find
\be\label{AH0_static}
\begin{split}
\sigma^{{\rm AH},0}_{yx}=\frac{e^2\Delta}{2\pi}\left(\frac{1}{\omega_{+}}-\frac{1}{\omega_{-}}\right)
=\frac{e^2}{4\pi}\left(\frac{\delta { \alpha}}{1+\delta^2}\right).
\end{split}
\ee
Here, $\delta \equiv\Delta/\sqrt{2m \alpha_{\rm R}^2 \varepsilon_{\rm F}}$, and $ { \alpha}\equiv \sqrt{2 m \alpha^2_{\rm R}/\varepsilon_{\rm F}}$, are the dimensionless gap and spin-orbit coupling parameters, respectively. 
Note that the order of limits is crucial here~\cite{GV_book}. The static (i.e., $\omega\to 0$) limit, through this paper, is performed after the clean system (i.e., $\omega\tau\to \infty$) limit is taken; therefore, we only have the intrinsic contribution to the conductivity.
The two-band regime criteria, i.e., $\Delta<\varepsilon_{\rm F}$ corresponds to $\delta<1/\alpha$.
The above expression for the anomalous Hall conductivity is identical to the results of Ref.~\cite{Nunner2007} in the two-band regime and vanishes for $\Delta\to 0$, as expected.

As for the noninteracting spin Hall conductivity, from the analytical forms of the non-interacting spin-spin response functions, we find~\cite{Supplemental}
\be\label{S-H-C-B}
\begin{split}
\sigma^{{\rm SH},0}_{yx}(\omega)&=\frac{e}{4m }\chi^{0}_{yy}(\omega)+\frac{ie\Delta}{2m \omega}  \chi^{0}_{yx}(\omega) \\
&=-\frac{e}{8\pi}\left[1+\left(\frac{\omega^2-4\Delta^2}{8m\alpha_{\rm R}^2 \omega  }\right)L(\omega)\right].
\end{split}
\ee
In zero gap limit, this reproduces the results of  Ref.~\cite{Berg2011}.
Furthermore, using the low-frequency expansion of $L(\omega)$ on the right-hand-side of Eq.~\eqref{S-H-C-B}, the results of Refs.~\cite{Marinescu2007,Hankiewicz2008} for the static spin Hall conductivity are recovered~\cite{Supplemental}
\be\label{St-S-H-C-Non}
\begin{split}
\sigma^{{\rm SH},0}_{yx}&=-\frac{e}{8\pi }\left[1-\frac{\Delta^2}{m\alpha_{\rm R}^2}\left(\frac{\omega_{-}-\omega_{+}}{\omega_{+}\omega_{-}}\right)\right]\\
&=-\frac{e}{8\pi}\left(\frac{1}{1+\delta^2}\right).
\end{split}
\ee
This expression reduces to the well-known universal value: $-e/(8\pi)$ in the $\Delta\to 0$ limit.

%%%%%%%%%%%%%%%%%%%%%%%%%%%%%%
\emph{Interacting spin-spin response functions.}--
Generally, we can express the interacting density and spin response functions in terms of the noninteracting ones in a random-phase-approximation (RPA) like expression~\cite{GV_book}
\be\label{Int-chi}
\chi(\qv,\omega)=\left[1-\chi^{0}(\qv,\omega)W(q,\omega)\right]^{-1}\chi^{0}(\qv,\omega).
\ee
Here, $\chi(q,\omega)$ and $\chi^0(q,\omega)$ are the $4\times 4$ matrices of the interacting and noninteracting spin-density linear response functions, whose elements are $\chi_{\alpha\beta}(q,\omega)$ and $\chi_{\alpha\beta}^0(q,\omega)$ where $\alpha,\beta=0,x,y,z$ with the $0$ index referring to the particle density. 
The exact effective $4\times 4$ interaction matrix $W(q,\omega)$ is in principle unknown. 
The celebrated RPA replaces it with the bare interaction
$W_{\alpha\beta}(q,\omega)=\delta_{\alpha,0}\delta_{\beta,0}v(q)$. Therefore within the RPA, spin responses are not modified by particle-particle interactions, and one recovers the noninteracting results for the anomalous and spin Hall conductivities~\cite{Amit2011,Mir}.

Several attempts have been made to go beyond the RPA~\cite{Amit2011, Maslov2015, Maslov2016,Maslov2017,Mir}. In this paper, we follow the procedure outlined in Ref.~\cite{Mir}
which extends the Hubbard local field factor concept~\cite{mahan_book,GV_book} for spin-orbit coupled systems. 
In this approximation, the effective interaction reads
\be\label{eq:effective-interaction}
W_{\alpha\beta}(q)=v(q)\delta_{\alpha,0}\delta_{\beta,0}-\frac{1}{2}v_{\rm H}(q)\delta_{\alpha,\beta},
\ee
where $v_{\rm H}(q)=v(\sqrt{k^{2}_{\rm F}+q^2})$~\cite{GV_book} is the screened Coulomb potential. 
In the long wavelength limit, this approximation becomes equivalent to the summation of ladder diagrams to infinite order with an ultrashort screened interaction, i.e $v_{\rm H}(q= 0) \approx U$~\cite{Maslov2015,Maslov2016,Maslov2017}.
Notice that our system in the two-band regime has two Fermi wave vectors. However, in the high Fermi energy, or small Zeeman energy and weak SOC limit, we can take $k_{\rm F}=\sqrt{2\pi n}$ as the average Fermi wave vector of the system, where $n$ is the electron density.

%%%%%%%%%%%%%%%%%%%%
\emph{Interacting anomalous and spin Hall conductivities.}--
Using the noninteracting spin-spin response functions in the long-wavelength limit~\cite{Supplemental} and the effective interaction with the Hubbard local field correction, we can find the interacting anomalous Hall and spin Hall conductivities. 

We begin with the anomalous Hall conductivity, which is proportional to the transverse in-plane spin susceptibility. Upon the substitution of the analytical form of $\chi_{yx}(\omega)$~\cite{Supplemental} in Eq.~\eqref{A-H-C-IN-B}, we find
\be\label{Dy-A-H-C}
\sigma^{\rm AH}_{yx}(\omega)
=\frac{\sigma^{\rm AH,0}_{yx}(\omega)}{\Gamma(\omega,U)}.
\ee
Here, the noninteracting anomalous Hall conductivity $\sigma^{{\rm AH},0}_{yx}(\omega)$ is as given in Eq. \eqref{A-H-C-Non}, and
\be\label{Gamma}
\Gamma(\omega,U)=\left[1+\frac{U}{2}\chi^{0}_{yy}(\omega)\right]^2+\left[\frac{U}{2}\chi^{0}_{xy}(\omega)\right]^2,
\ee 
where the  interaction strength is $U=v_{\rm H}(q=0)=2\pi e^2/k_{\rm F}$, and $\chi^{0}_{yy}(\omega)$ and $\chi^{0}_{xy}(\omega)$ are the noninteracting spin-spin linear response functions.
The static anomalous Hall conductivity is obtained from the $\omega \rightarrow 0 $ limit of Eq.~\eqref{Dy-A-H-C}
\be\label{St-A-H-C}
\begin{split}
\sigma^{\rm AH}_{yx}
=\frac{e^2}{4\pi}\frac{{\alpha}\delta \left(1+\delta^2\right) }
{\left[\left(1+\delta^2\right)-{u}/2\left(1+2 \delta^2\right)\right]^2},
\end{split}
\ee
where $u=mU/(2\pi)=me^2/k_{\rm F}$ is the dimensionless interaction strength. 

For the interacting dynamical spin Hall conductivity, substituting $\chi_{yy}(\omega)$ and $\chi_{xy}(\omega)$
 in Eq.~\eqref{S-H-C-IN-B}, we obtain
\be\label{Dy-S-H-C}
\sigma^{\rm SH}_{yx}(\omega)=\frac{\sigma^{{\rm SH},0}_{yx}(\omega)}{\Gamma(\omega,U)}+
\left(\frac{eU}{8m}\right)\frac{[\chi^{0}_{yy}(\omega)]^2+[\chi^{0}_{xy}(\omega)]^2}{\Gamma(\omega,U)}.
\ee
Here, the noninteracting dynamical spin Hall conductivity $\sigma^{{\rm SH},0}_{yx}(\omega)$ is given in Eq.~\eqref{S-H-C-B}.
The static spin Hall conductivity is obtained from the $\omega \rightarrow 0 $ limit of the dynamical one. 
Upon inserting $\chi^{0}_{xy}(\omega = 0)$ and $\chi^{0}_{yy}(\omega = 0)$~\cite{Supplemental} in Eq.~\eqref{Dy-S-H-C}, we have
\be\label{eq:static-spin-hall-int}
\begin{split}
\sigma^{\rm SH}_{yx}=-\frac{e}{8\pi}
\frac{\left(1+\delta^2\right)-{u}/2\left(1+2 \delta^2\right)^2 }
{\left[\left(1+\delta^2\right)-{u}/2\left(1+2 \delta^2\right)\right]^2}.
\end{split}
\ee
In the gapless limit, this simplifies to~\cite{Mir}
\be
\left.\sigma^{{\rm SH}}_{yx}\right|_{\Delta=0}=-\frac{e}{8\pi}\frac{1}{1-{ u}/2},
\ee
which coincides with the result obtained within the time-dependent Hartree-Fock approximation~\cite{Amit2011}.
A fascinating prediction of Eq.~\eqref{eq:static-spin-hall-int} is the sign reversal of the spin-Hall conductivity for 
$u>2(1+\delta^2)/(1+2 \delta^2)^2$. The interaction strength for this sign reversal can be accessibly small for large exchange fields. 

%%%%%%%%%%%%%%%%%%%%%%%%%%%%%%%%%%%%
\emph{Results and discussion.}--
Now we discuss the results obtained from the the many-body correction on the intrinsic anomalous Hall and spin Hall conductivities in the static limit and within the Hubbard approximation for the local field factor. We should emphasize once more that the static limits are taken after the clean system limits.
%%%%%%%%%%%Fig.1%%%%%%%%%%%%%%%%%%
\begin{figure}
\begin{tabular}{cc}
\hspace{-1.2em}\includegraphics[width=0.55\linewidth]{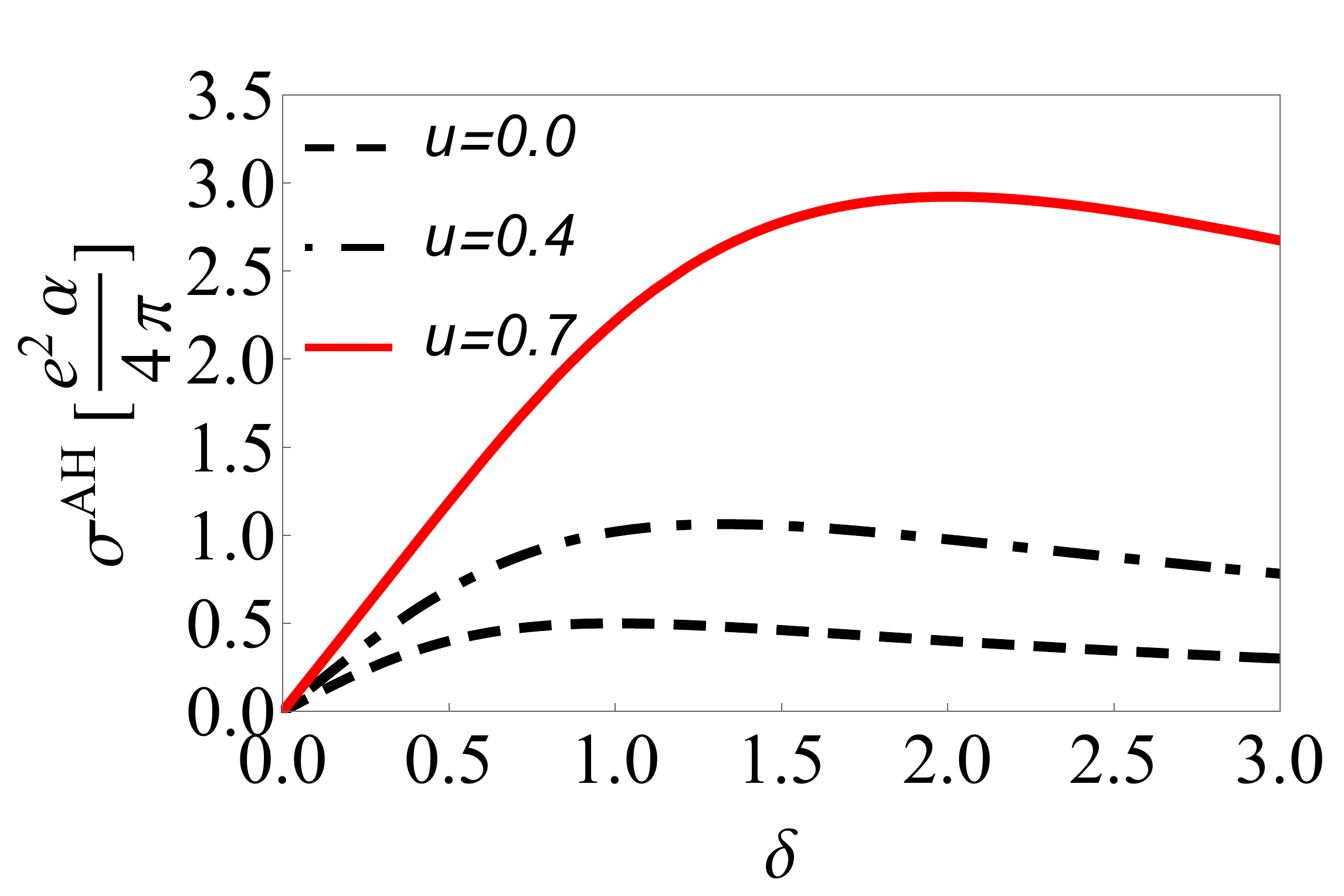}
\hspace{-1.5em}\includegraphics[width=0.55\linewidth]{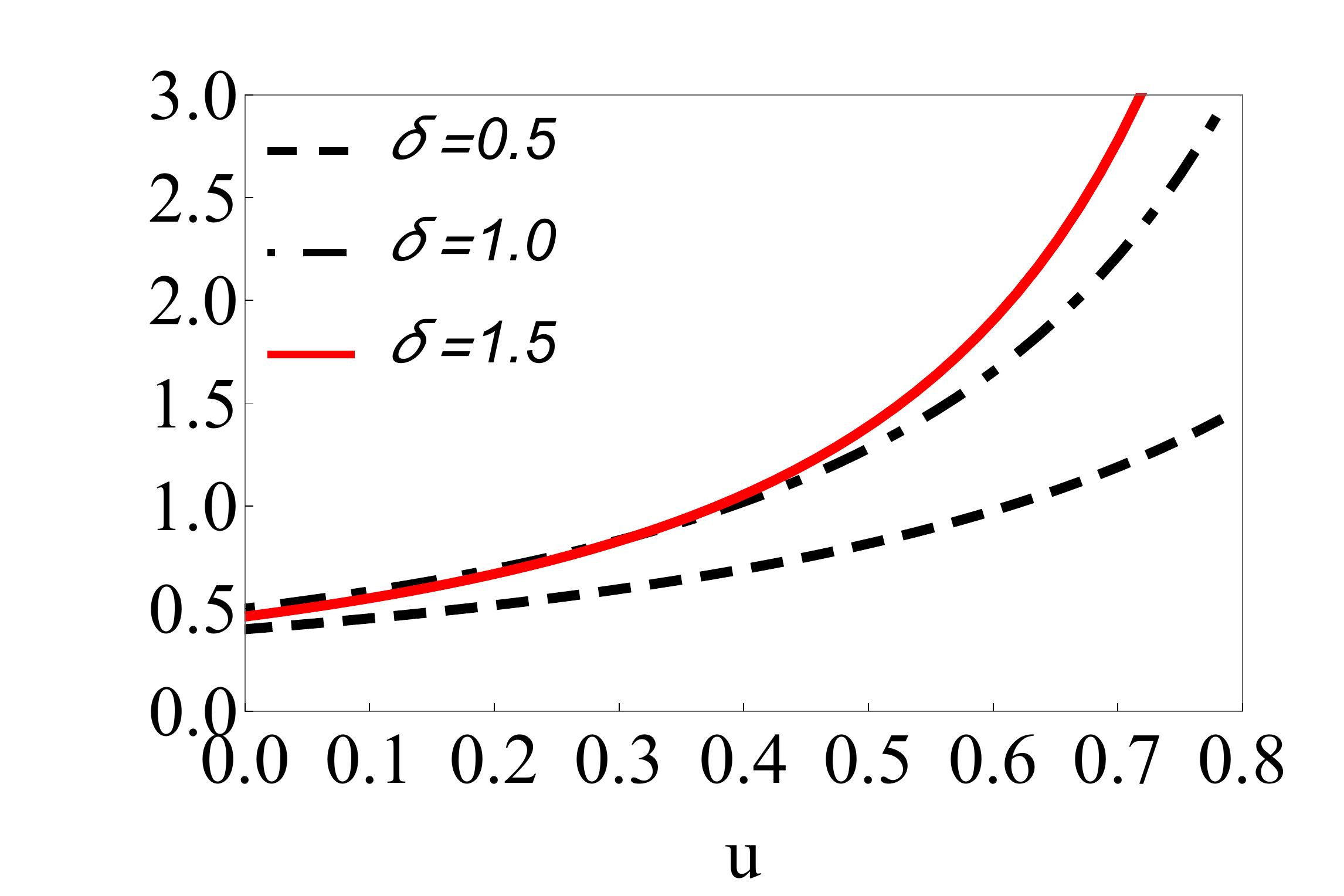}
\end{tabular}
\caption{Interacting static anomalous Hall conductivity [in units of $\alpha e^2/(4\pi)$] versus the dimensionless gap parameter  $\delta$ for different values of the interaction strength ${u}$  (left), and versus $u$ for different values of 
 $\delta$ (right).
\label{fig:St-AH}}
\end{figure}

The dependence of the static anomalous Hall conductivity on the exchange field $\delta =\Delta/\sqrt{2m \alpha_{\rm R}^2 \varepsilon_{\rm F}}$ and the electron-electron interaction strength $u=m U/(2\pi)$ is illustrated in Fig.~\ref{fig:St-AH}. 
We observe that the electron-electron interaction enhances the intrinsic anomalous Hall conductivity, and its effect is more pronounced in systems with large exchange fields or weak spin-orbit couplings. 
Moreover, the static anomalous Hall conductivity non-monotonically depends on the exchange parameter.
  
\begin{figure}
\centering
\begin{tabular}{cc}
\hspace{-1.2em}\includegraphics[width=0.55\linewidth]{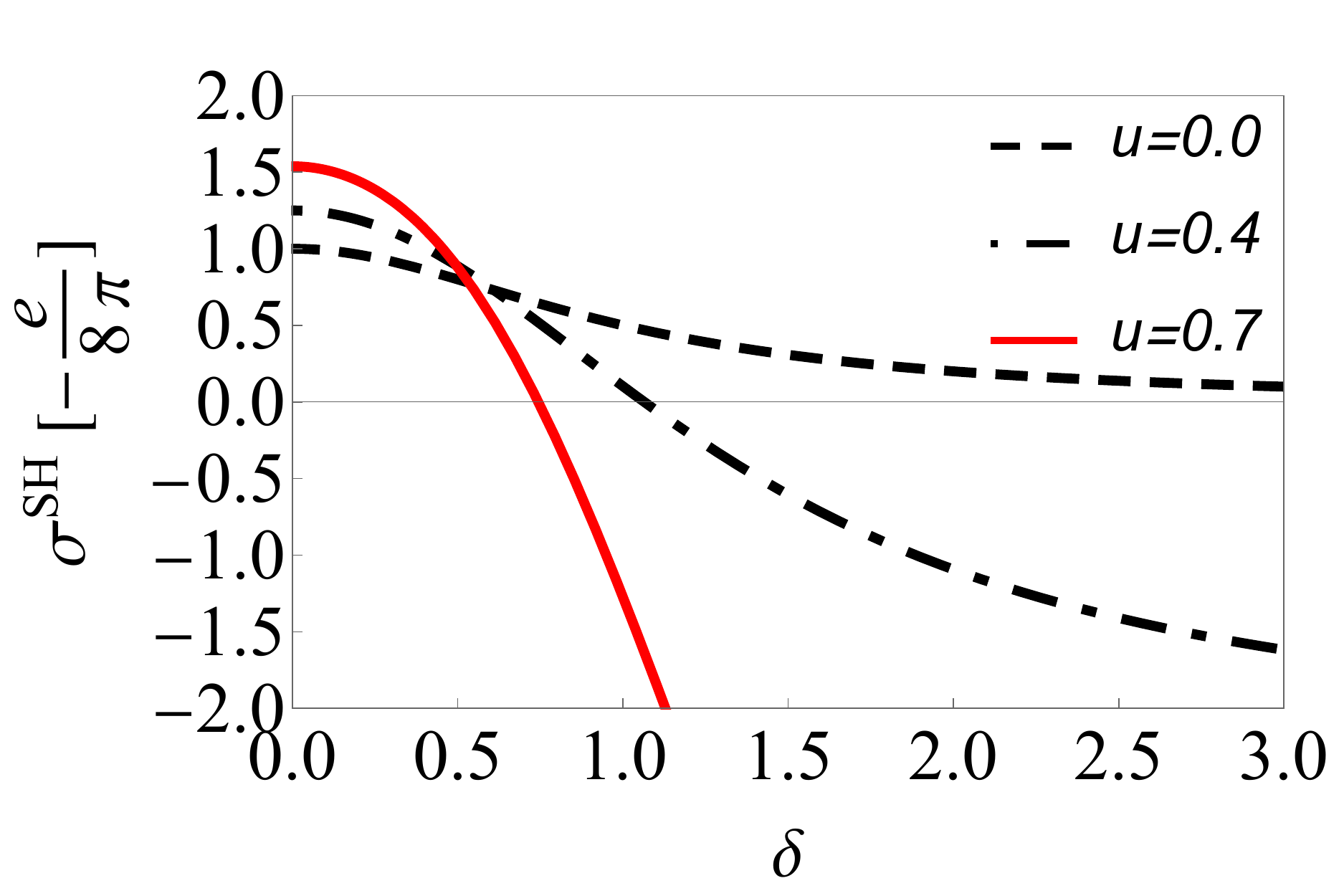}
\hspace{-1.5em}\includegraphics[width=0.55\linewidth]{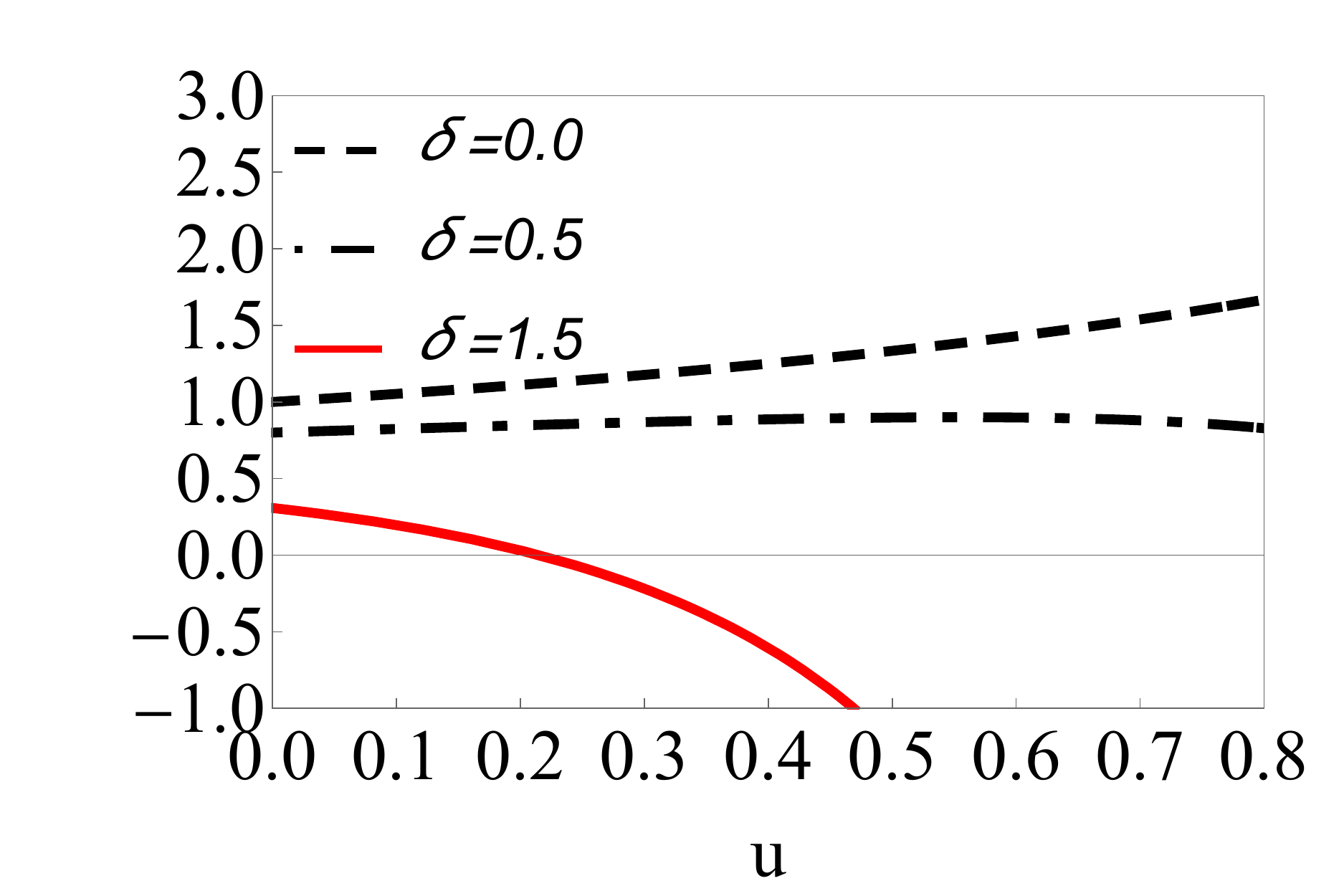}
\end{tabular}
\caption{Interacting static spin Hall conductivity [in units of $-e/(8\pi)$] versus gap parameter $\delta$ for different values of the interaction strength $u$ (left), and versus $u$ for several values of $\delta$  (right).}
\label{fig:St-SH}
\end{figure}

Similarly, Fig.~\ref{fig:St-SH} shows the effects of exchange field and electron-electron interaction on the static spin Hall conductivity.
The interaction strength dependence is generally non-monotonic. In particular, at large exchange fields (or weak spin-orbit couplings), strong electron-electron interaction strength leads to the sign reversal of the static spin Hall conductivity.
%%%%%%%%Fig.2%%%%%%%%%%%%

\begin{figure}
\centering
\begin{tabular}{cc}
\hspace{-1.2em}
\includegraphics[width=0.53\linewidth]{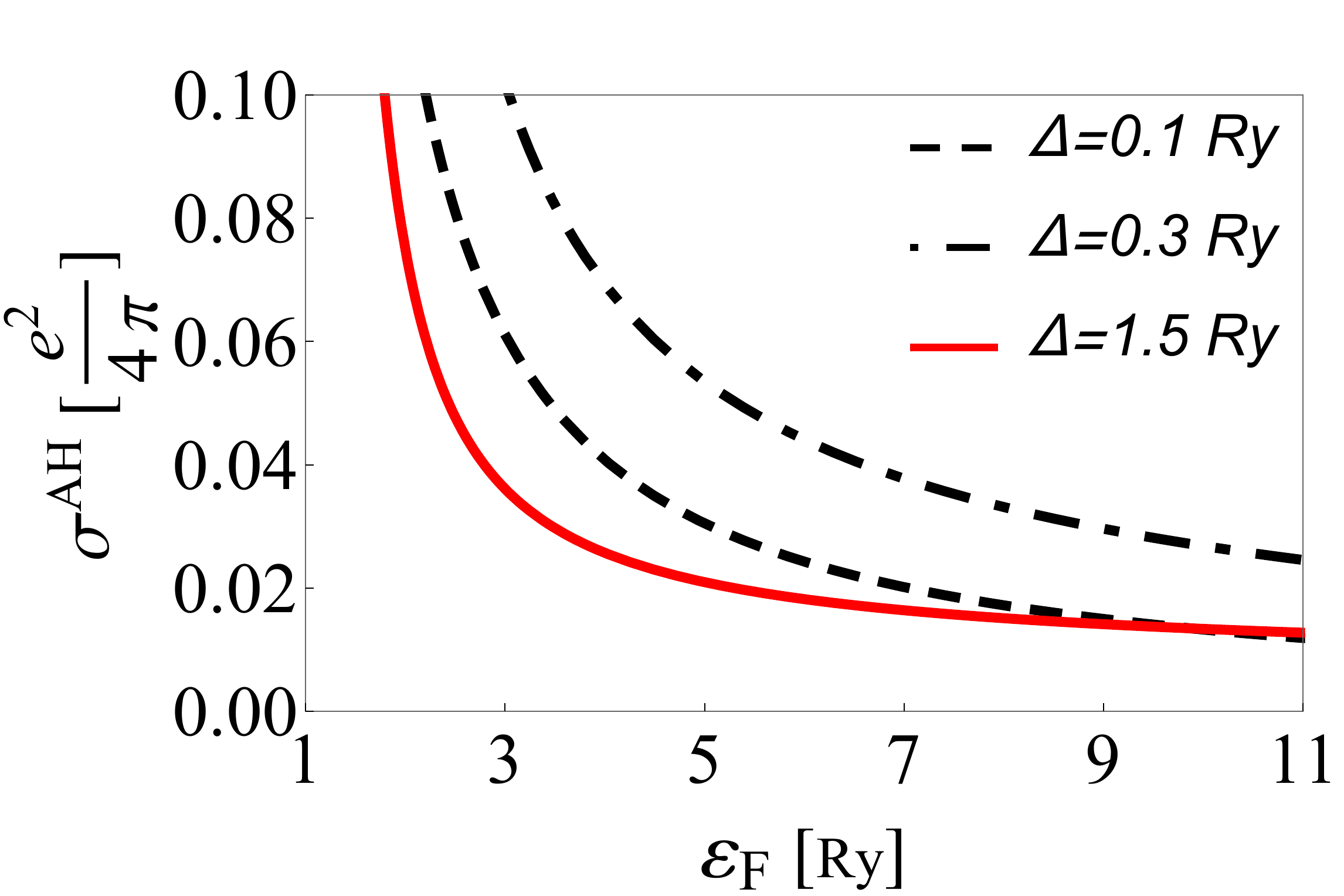}
\hspace{-1.em}
\includegraphics[width=0.53\linewidth]{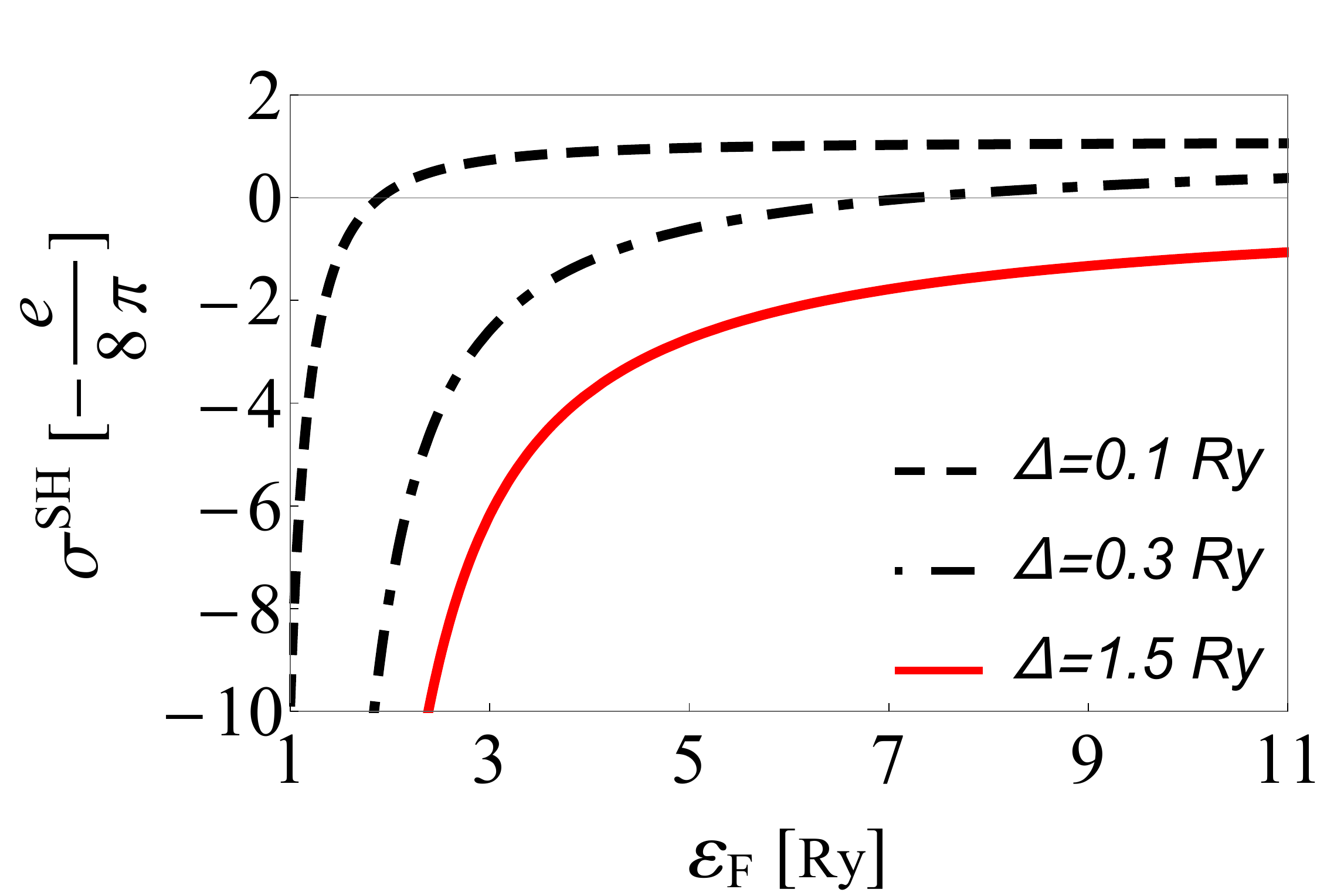}
\end{tabular}
\caption{Interacting static anomalous Hall (left) and spin Hall (right) conductivities versus Fermi energy $\varepsilon_{\rm F}$ for different values of $\Delta$. Here we have used $\sqrt{2m \alpha^2_{\rm R}/{\rm Ry}}=0.1$, for the spin-orbit coupling strength.}
\label{fig:St-Fermi}
\end{figure}

In the laboratory, it is most convenient to tune the chemical potential of the two-dimensional electron gas. Our dimensionless parameters $u$, $\delta$, and $\alpha$, all depend on the Fermi energy. 
To reveal the Fermi energy dependence of the anomalous and spin Hall conductivities, we should notice that $\varepsilon_{\rm F}=n \pi/m-m \alpha^2_{\rm R}$, in the two band regime. Then, we find  $u=\sqrt{{\rm Ry}/\left(\varepsilon_{\rm F}+ m \alpha^2_{\rm R}\right)}$, where ${\rm Ry}=e^2/(2 a_{\rm B})$ is the effective Rydberg energy, with $a_{\rm B}=1/(me^2)$ being the effective Bohr radius~\cite{GV_book}.
In Fig.~\ref{fig:St-Fermi}, we illustrate the Fermi energy dependence of the static anomalous Hall conductivity and spin Hall conductivity.
Similar trend to Figs.~\ref{fig:St-AH} and~\ref{fig:St-SH} is evident here too. At low densities, which corresponds to the strong interaction regime, the anomalous Hall conductivity is enhanced, and the sign of the spin Hall conductivity is reversed.
%%%%%%%%Fig.3%%%%%%%%%%%%

%%%%%%%%%%%%%%%%%%%%%%%%%%
\emph{Experimental feasibility.}--
To check whether what we predict here could also be experimentally verified, we take the InAs quantum well, which hosts two-dimensional electron gas with Rashba spin-orbit coupling, as a realistic material example. The effective mass of electrons in InAs quantum well is $m \approx 0.023\,m_{e}$, where $m_{e}$ is the bare electron mass. Its high-frequency dielectric constant is $\epsilon \approx 15$. These values correspond to $a_{\rm B} \approx 345\, \AA$ and ${\rm Ry}\approx 1.4\, {\rm meV}$, for the effective Bohr radius and Rydberg energy, respectively. 
The Rashba spin-orbit coupling constant in InAs varies in the range $\alpha_{\rm R} \approx (10-60)\, {\rm meV}\, \AA$~\cite{Amit2011}. 
An electronic density of $n \approx (2-8) \times 10^{10}\, {\rm cm}^{-2}$, and moderate exchange energy of $\Delta\approx ~(0.1-0.4) \, {\rm meV}$, would be suitable to observe the sign reversal of the spin Hall conductivity and significant enhancement of the anomalous Hall conductivity.

Notice that our results are obtained in the zero temperature and clean limit. 
As the typical energy scales in realistic two-dimensional quantum wells with SOC are of the order of milli-electron volts, very low temperatures of a few Kelvins are required in the experiments. 
The inclusion of disorder complicates the problem as the extrinsic mechanisms for the anomalous and spin Hall conductivities compete with the intrinsic ones. However, high-frequency measurements are powerful techniques to probe intrinsic responses.

%%%%%%%%%%%%%%%%%%%%%%%%%%%%%%%%%
\emph{Acknowledgements.}--
We thank Dimitrie Culcer and Giovanni Vignale for their insightful comments.
The work in Zanjan is supported by the Iranian Science Elites Federation (ISEF) and the Research Council of the Institute for Advanced Studies in Basic Sciences (IASBS). M. M. acknowledges the kind hospitality of the Department of Physics at IASBS during the final stages of preparing this work.

%%%%%%%%%%%%%%%%%%%%%%%%%%%%%%%%%

%%%%%%%%%%%%%%%%%%%%%
\newpage
%%%%%%%%%% Merge with supplemental materials %%%%%%%%%%
\pagebreak
\clearpage
\widetext
\begin{center}
\textbf{\large Supplemental Materials for "Many-body correction on the intrinsic anomalous and spin Hall conductivities"}

\end{center}
%%%%%%%%%% Merge with supplemental materials %%%%%%%%%%
%%%%%%%%%% Prefix a "S" to all equations, figures, tables and reset the counter %%%%%%%%%%
\setcounter{equation}{0}
\setcounter{figure}{0}
\setcounter{table}{0}
\setcounter{page}{1}
\makeatletter
\renewcommand{\theequation}{S\arabic{equation}}
\renewcommand{\thefigure}{S\arabic{figure}}
\renewcommand{\bibnumfmt}[1]{[S#1]}
\renewcommand{\citenumfont}[1]{S#1}
%%%%%%%%%% Prefix a "S" to all equations, figures, tables and reset the counter %%%%%%%%%%

%\begin{appnedix}
%\begin{widetext}
%\appendix

\section{\label{app:formfactor} Form factors of the spin-density operators for magnetic 2D Rashba electron gas}
We can write the spin-density operators in the band basis as
\be
\h{S}^{i}_\qv=\sum_{\kv,\mu\nu} F^{i}_{\mu,\nu}(\kv-\frac{\qv}{2};\kv+\frac{\qv}{2})\hdg{c}_{\kv-\frac{\qv}{2},\mu}\h{c}_{\kv+\frac{\qv}{2},\nu},
\ee
with the form-factor matrices being defined as
\be
F^{i}(\kv-\frac{\qv}{2};\kv+\frac{\qv}{2}) = U^\dagger(\kv-\frac{\qv}{2}) \tau^i U(\kv+\frac{\qv}{2}).
\ee
Here, $\tau^0=\mathbb{I}$, $\tau^{i\neq 0}$ are $2\times2$ Pauli matrices, and $U(\kv)$ is the unitary  matrix which diagonalizes the single-particle part of the Hamiltonian [i.e., Eq.~\eqref{hamilH_{0}}] and is obtained from the eigenvectors Eq.~\eqref{vectors}
\be
U(\bf{k}) =\left(\begin{array}{cc} 
\sin(\varphi_{k,+}/2)&\sin(\varphi_{k,-}/2)\\ 
-i e^{i\theta_{k}}\cos(\varphi_{k,+}/2) &i e^{i\theta_{k}} \cos(\varphi_{k,-}/2) 
\end{array}\right) , 
\ee
therefore different form factors at $q=0$ read
\be
\begin{split}
F^{0}_{\mu,\nu}(\kv,\kv)&=\sin(\frac{\varphi_{\kv,\mu}}{2})\sin(\frac{\varphi_{\kv,\nu}}{2})+\mu \nu \cos(\frac{\varphi_{\kv,\mu}}{2}) \cos(\frac{\varphi_{\kv,\nu}}{2}),\\
F^{x}_{\mu,\nu}(\kv,\kv)&=-i\nu e^{i\theta_{\kv}}\sin(\frac{\varphi_{\kv,\mu}}{2})\cos(\frac{\varphi_{\kv,\nu}}{2})+i\mu  e^{-i\theta_{\kv}}\cos(\frac{\varphi_{\kv,\mu}}{2}) \sin(\frac{\varphi_{\kv,\nu}}{2}),\\
F^{y}_{\mu,\nu}(\kv,\kv)&=-\mu e^{i\theta_{\kv}}\cos(\frac{\varphi_{\kv,\mu}}{2})\sin(\frac{\varphi_{\kv,\nu}}{2})-\nu  e^{-i\theta_{\kv}}\sin(\frac{\varphi_{\kv,\mu}}{2}) \cos(\frac{\varphi_{\kv,\nu}}{2}),\\
F^{z}_{\mu,\nu}(\kv,\kv)&=\sin(\frac{\varphi_{\kv,\mu}}{2})\sin(\frac{\varphi_{\kv,\nu}}{2})-\mu \nu \cos(\frac{\varphi_{\kv,\mu}}{2}) \cos(\frac{\varphi_{\kv,\nu}}{2}),
\end{split}
\ee
with $\mu,\nu=\pm 1$.

\section{\label{app:Non-Int-spin-density}Noninteracting dynamical spin-density response functions of  magnetic 2D Rashba electron gas}
The different components of the noninteracting spin-density response functions are defined as
\be
\chi^{0}_{\alpha\beta}(\qv,\omega)=\sum_{\kv,\mu,\nu}\frac{n_{\kv_1,\mu}-n_{\kv_2,\nu}}{\hbar \omega+\varepsilon_{\kv_1,\mu}-\varepsilon_{\kv_2,\nu}}
F^{\alpha}_{\mu,\nu}(\kv_1,\kv_2)F^{\beta}_{\nu,\mu}(\kv_2,\kv_1),
\ee
where $\kv_1=\kv-\qv/2$, $\kv_2=\kv+\qv/2$, $n_{k,\mu}$ is the Fermi-Dirac distribution function, and $F^{i}_{\mu,\nu}$ is the form factor introduced in Sec.~\ref{app:formfactor}.
In the long-wavelength, i.e., $ \qv\to 0$ limit, after some algebra, we obtain
%
%
%
%\be
%\begin{split}
%\chi^{0}_{\alpha,\beta}(0,\omega)=\sum_{\kv}\frac{n_{\kv,+}-n_{\kv,+}}{\hbar \omega+\varepsilon_{\kv,+}-\varepsilon_{\kv,+}}
%F^{\alpha}_{+,+}(\kv,\kv)F^{\beta}_{+,+}(\kv,\kv)
%+\sum_{\kv}\frac{n_{\kv,+}-n_{\kv,-}}{\hbar \omega+\varepsilon_{\kv,+}-\varepsilon_{\kv,-}}
%F^{\alpha}_{+,-}(\kv,\kv)F^{\beta}_{-,+}(\kv,\kv)
%\\
%\sum_{\kv}\frac{n_{\kv,-}-n_{\kv,+}}{\hbar \omega+\varepsilon_{\kv,-}-\varepsilon_{\kv,+}}
%F^{\alpha}_{-,+}(\kv,\kv)F^{\beta}_{+,-}(\kv,\kv)
%+\sum_{\kv}\frac{n_{\kv,-}-n_{\kv,-}}{\hbar \omega+\varepsilon_{\kv,-}-\varepsilon_{\kv,-}}
%F^{\alpha}_{-,-}(\kv,\kv)F^{\beta}_{-,-}(\kv,\kv)
%\end{split}
%\ee
%
%and after some calculation, we obtain
\be
\chi^{0}_{\alpha\beta}(0,\omega)=\sum_{\kv}\left[
\frac{F^{\alpha}_{-,+}(\kv,\kv)F^{\beta}_{+,-}(\kv,\kv)}{\hbar \omega+\varepsilon_{\kv,-}-\varepsilon_{\kv,+}}
-\frac{F^{\alpha}_{+,-}(\kv,\kv)F^{\beta}_{-,+}(\kv,\kv)}{\hbar \omega+\varepsilon_{\kv,+}-\varepsilon_{\kv,-}}
\right](n_{\kv,-}-n_{\kv,+}).
\ee
In this $q=0$ limit, the non-vanishing elements of the spin-density response matrix are 
\be\label{eq:chi0_q0}
\begin{split}
\chi^{0}_{xx}(0,\omega)&=\chi^{0}_{yy}(0,\omega)=-\frac{1}{2\pi \alpha_{\rm R}^2}\left[\frac{\omega_{-}-\omega_{+}}{4}+\frac{(\omega^2+4\Delta^2)}{8\omega}L(\omega)\right],\\
\chi^{0}_{xy}(0,\omega)&=-\chi^{0}_{yx}(0,\omega)=\frac{i\Delta}{4 \pi \alpha_{\rm R}^2}L(\omega),\\
\chi^{0}_{zz}(0,\omega)&=-\frac{1}{\pi \alpha_{\rm R}^2}\left[\frac{\omega_{-}-\omega_{+}}{4}+\frac{(\omega^2-4\Delta^2)}{8\omega}L(\omega)\right],
\end{split}
\ee
where $L(\omega)$ and $\omega_\pm$ are as defined in the main text.
Furthermore, it is straightforward to verify that $\chi_{0x}$,$\chi_{x0}$, $\chi_{0y}$,$\chi_{y0}$, $\chi_{xz}$,$\chi_{zx}$, $\chi_{yz}$,$\chi_{zy}$ linearly, and  $\chi_{00}$, $\chi_{0z}$, $\chi_{z0}$ quadratically vanish at $q\to0$ limit.
To obtain the static limits of the Hall conductivities, it is necessary to find the  $\omega \rightarrow 0$ limit of the dynamical response functions, therefore replacing
the $\omega \rightarrow 0$ limit of the $L(\omega)$, i.e.,
\be
L(\omega \rightarrow 0)\approx 2\left(\frac{\omega_{-}-\omega_{+}}{\omega_{-}\omega_{+}}\right)\omega,
\ee
in Eq.~\eqref{eq:chi0_q0}, we obtain
\be
\begin{split}
\chi^{0}_{xx}(0,\omega \rightarrow 0)&=\chi_{yy}(0,\omega \rightarrow 0)=-\frac{1}{2\pi \alpha_{\rm R}^2}\left[\frac{\omega_{-}-\omega_{+}}{4}+\Delta^2\left(\frac{\omega_{-}-\omega_{+}}{\omega_{-}\omega_{+}}\right)\right],\\
\chi^{0}_{xy}(0,\omega \rightarrow 0)&=-\chi_{yx}(0,\omega \rightarrow 0)=\frac{i\Delta \omega}{2 \pi \alpha_{\rm R}^2}\left(\frac{\omega_{-}-\omega_{+}}{\omega_{-}\omega_{+}}\right),\\
\chi^{0}_{zz}(0,\omega \rightarrow 0)&=-\frac{1}{\pi \alpha_{\rm R}^2}\left[\frac{\omega_{-}-\omega_{+}}{4}-\Delta^2\left(\frac{\omega_{-}-\omega_{+}}{\omega_{-}\omega_{+}}\right)\right].
\end{split}
\ee

%%%%%%%%%%%%%%%%%%%%%%%%%
\section{\label{app:Int-spin-density}Interacting dynamical spin-density responses of magnetic Rashba 2DEG}
Elements of the interacting spin-density response matrix are obtained from Eq.~\eqref{Int-chi}
\be
\chi(\qv,\omega)=[1-\chi^{0}(\qv,\omega) W(q)]^{-1}\chi^{0}(\qv,\omega).
\ee
Here, $\chi^{0}(\qv,\omega)$ is the $4\times 4$ noninteracting spin-density response function matrix and $W(q)$ is the $4\times 4$ effective interaction matrix within the Hubbard approximation for the local field factor, which we have introduced in Eq.~\eqref{eq:effective-interaction}.
In the $\qv \rightarrow 0$ limit, using the results obtained in Sec.~\ref{app:Non-Int-spin-density}, and Eq.~\eqref{eq:effective-interaction} for the effective interaction, we find
\be
\begin{split}
\chi_{xx}(\omega)&=\chi_{yy}(\omega)=\frac{[1+\frac{U}{2}\chi^{0}_{yy}(\omega)]\chi^{0}_{yy}(\omega)+\frac{U}{2}[\chi^{0}_{xy}(\omega)]^2}{\Gamma[\omega,U]},\\
\chi_{xy}(\omega)&=-\chi_{yx}(\omega)=\frac{\chi^{0}_{xy}}{\Gamma[\omega,U]},\\
\chi_{zz}(\omega)&=\frac{\chi^{0}_{zz}(0,\omega)}{1+\frac{U}{2}\chi^{0}_{zz}(0,\omega)},
\end{split}
\ee
where 
\be
\Gamma(\omega,U)=\left[1+\frac{U}{2}\chi^{0}_{yy}(\omega)\right]^2+\left[\frac{U}{2}\chi^{0}_{xy}(\omega)\right]^2.
\ee

\section{\label{app:Non-Int-dynamical-S-H-C}
Noninteracting dynamical spin Hall conductivity of non-magnetic Rashba gas}

The noninteracting dynamical spin Hall conductivity in zero exchange field  $(\Delta \rightarrow 0)$ limit is  obtained from Eq.~\eqref{S-H-C-B}
\be\label{eq:sigma_sh_d0}
\left.\sigma^{\rm SH,0}(\omega)\right|_{\Delta=0}=
-\frac{e}{8\pi m\alpha^2_{\rm R}}\left\{\frac{\omega_{-}-\omega_{+}}{4}+\frac{\omega}{8}\log\left[\frac{(\omega-\omega_{-}+i0^+)(\omega+\omega_{+}+i0^+)}{(\omega-\omega_{+}+i0^+)(\omega+\omega_{-}+i0^+)}\right]\right\},
\ee
where $\omega_{\pm}$ are introduced in the main text. Putting $\Delta = 0$, we find
\be
\omega_{\pm}=2\sqrt{2m\alpha^{2}_{\rm R} \varepsilon_{\rm F}+(m\alpha_{\rm R}^2)^2} \mp 2m\alpha_{\rm R}^2.
\ee
Upon the substitution of $\omega_{-}-\omega_{+}=4m\alpha^2_{\rm R}$, and $\omega_{+}\omega_{-}=8\varepsilon_{\rm F} m\alpha_{\rm R}$ in Eq.~\eqref{eq:sigma_sh_d0}, for the real part of dynamical spin Hall conductivity, we get
\be\label{Real-sigma-1}
\left.
{\rm Re}~\sigma^{\rm SH,0}(\omega)\right|_{\Delta=0}=
-\frac{e}{8\pi }\left(1-\frac{\omega}{8m \alpha^2_{\rm R}}\log \left[\frac{\omega(\omega+4m\alpha^2_{\rm R})-8\varepsilon_{\rm F} m \alpha_{\rm R})}{\omega(\omega-4m\alpha^2_{\rm R})-8\varepsilon_{\rm F} m \alpha_{\rm R})}\right]\right),
\ee
which is the same as what was obtained in Ref.~\cite{S-Berg2011}.

%%%%%%%%%%%%%%%%%%%%%%%
\section{\label{app:Non-Int-static-S-H-C-B}Non-interacting static spin Hall conductivity at finite exchange field}
The static spin Hall conductivity could be obtained from the zero frequency limit of Eq.~\eqref{S-H-C-B}
\be
\begin{split}\label{static-SHC}
\sigma^{\rm SH,0}(\omega\rightarrow 0)&=\frac{e}{8\pi m \alpha^2_{\rm R}}\left[\frac{\omega_{+}-\omega_{-}}{4}-\Delta^2\left(\frac{\omega_{+}-\omega_{-}}{\omega_{+}\omega_{-}}\right)\right]\\
&=\frac{e}{8\pi m \alpha^2_{\rm R}}\left[\frac{\omega^{2}_{+}+4\Delta^2}{4\omega_{+}}-\frac{\omega^{2}_{-}+4\Delta^2}{4\omega_{-}}\right].
\end{split}
\ee
To compare our result with Ref.~\cite{S-Marinescu2007}, we rewrite $\omega_{\pm}$ as
\be\label{omega_pm-2}
\omega_{\pm}=2\sqrt{2m\alpha^2_{\rm R} \epsilon_{\rm F,\pm}+\Delta^2},
\ee
where $\epsilon_{\rm F, \pm}\equiv k^2_{\rm F, \pm}/2m$, with
\be
k_{\rm F,\pm}=
\sqrt{2(\varepsilon_{\rm F} m+m^2\alpha^2_{\rm R})\mp 2\sqrt{(m^2\alpha^2_{\rm R})^2+2\varepsilon_{\rm F} m(m^2\alpha^2_{\rm R})+m^2\Delta^2}}.
\ee
Now,  Eq.~\eqref{static-SHC} can be written as
\be
\sigma^{\rm SH,0}(\omega\rightarrow 0)=
\frac{e}{8\pi m \alpha^2_{\rm R}}\left[\frac{\Delta^2+m\alpha^2_{\rm R}\epsilon_{\rm F,+}}{\sqrt{\Delta^2+2m\alpha^2_{\rm R}\epsilon_{\rm F,+}}}-\frac{\Delta^2+m\alpha^2_{\rm R}\epsilon_{\rm F,-}}{\sqrt{\Delta^2+2m\alpha^2_{\rm R}\epsilon_{\rm F,-}}}\right],
\ee
which is identical to the noninteracting dynamical spin Hall conductivity reported by Moca and Marinescu in Ref.~\cite{S-Marinescu2007}.

%%%%%%%%%%%%%%%%%%%%%%%%%%%%%%%%%%%
\section{Dynamical anomalous Hall and spin Hall conductivities \label{app:dynamical}}
\begin{figure}
	\centering
	\begin{tabular}{cc}
		\hspace{-2em}\includegraphics[width=0.55\linewidth]{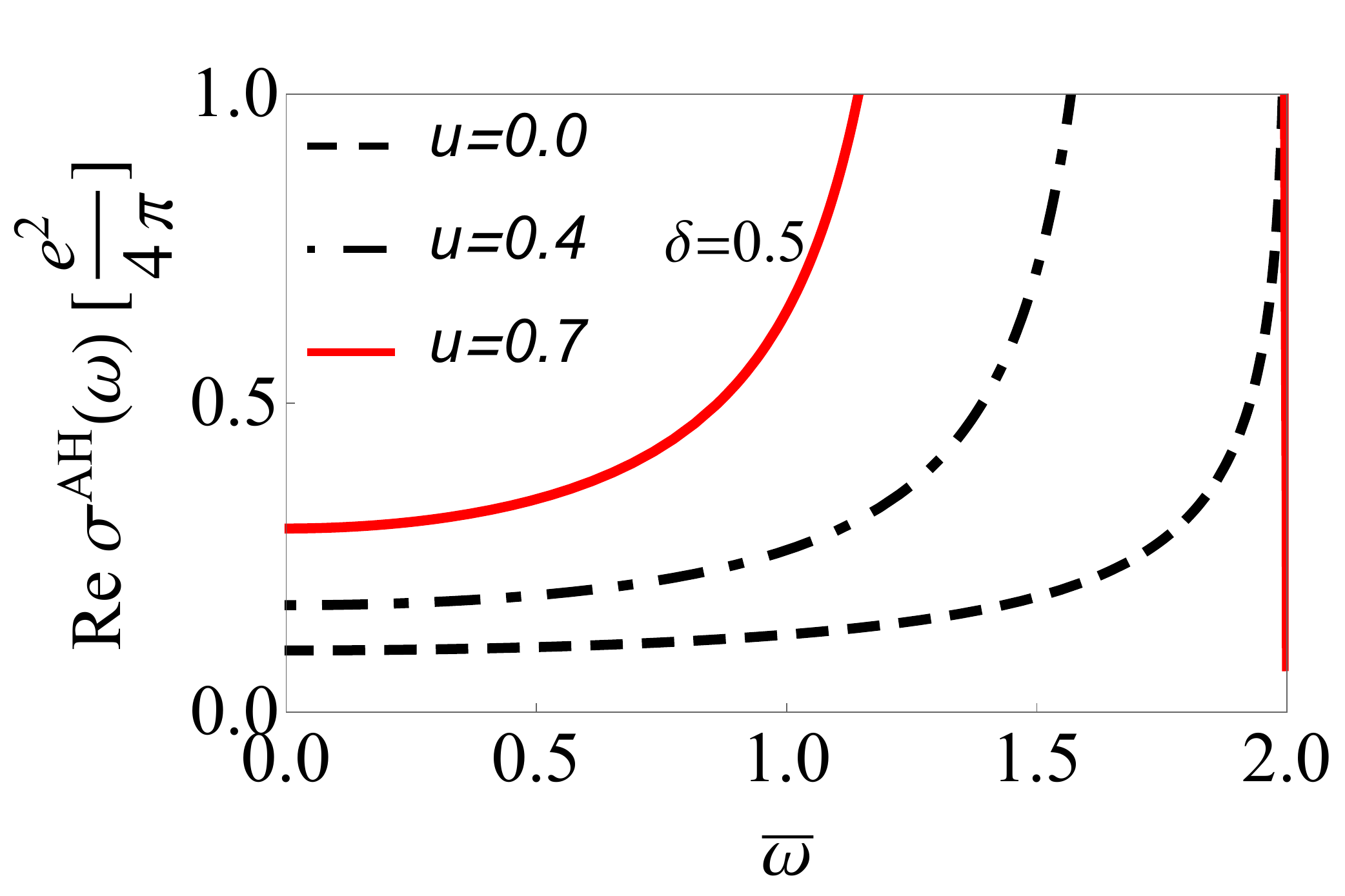} 		
		\hspace{-3.5em} \includegraphics[width=0.55\linewidth]{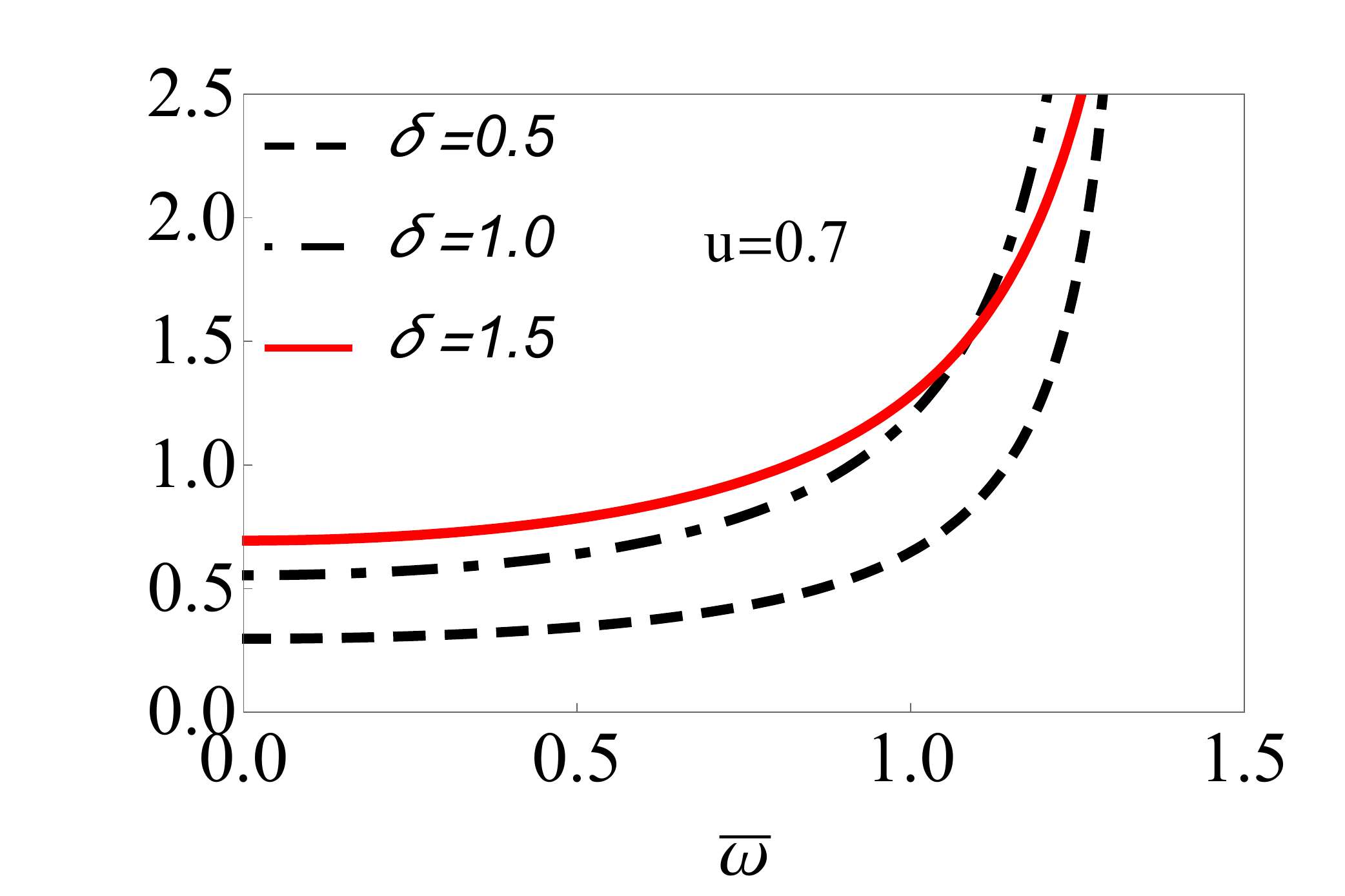}
	\end{tabular}
	\caption{Interacting dynamical anomalous Hall conductivity [in units of $e^2/(4\pi)$] versus dimensionless frequency $\bar{\omega}=\omega/\sqrt{2m\alpha^2_{\rm R} \epsilon_{\rm F}}$, for $\delta = 0.5$ and different values of the interaction 
		strength $u$ (left), and for $u = 0.7$  and different values of the exchange field
	 $\delta$ (right). We have used $\alpha=0.25$ for the dimensionless spin-orbit coupling strength in both panels. 
	\label{fig:Dy-AHC}}
\end{figure}

For the interacting anomalous Hall conductivity, substituting $\sigma^{\rm AH,0}_{yx}(\omega)$ from Eq.~\eqref{A-H-C-Non} and $\chi^{0}_{yy}(\omega)$ and $\chi^{0}_{xy}(\omega)$ form Appendix~\ref{app:Non-Int-spin-density} in Eq.~\eqref{Dy-A-H-C}, we find
\be
\sigma^{\rm AH}_{yx}({\omega})=\frac{e^2}{4\pi}\frac{ \frac{\delta}{\bar{\omega}}L({\omega}) }{ \left[1-\frac{u}{2}\left(1+\frac{\bar{\omega}^2+4\delta^2}{4\alpha \bar{\omega}}L({\omega})\right)\right]^2-\left[\frac{u \delta}{2 \alpha}  L({\omega})\right]^2 },
\ee
where $\bar{\omega}=\omega/\sqrt{2m\alpha^2_{\rm R} \epsilon_{\rm F}}$ is the dimensionless frequency.
In Fig.~\ref{fig:Dy-AHC}, we plot the real part of the interacting dynamical anomalous Hall conductivity versus $\bar{\omega}$, for different values of the interaction strength $u$, and exchange field $\delta$. 
%%%%%%%%%%%Fig.3%%%%%%%%%%%%%%%%%%

\begin{figure}
\centering
\begin{tabular}{cc}
\hspace{-2em}\includegraphics[width=0.55\linewidth]{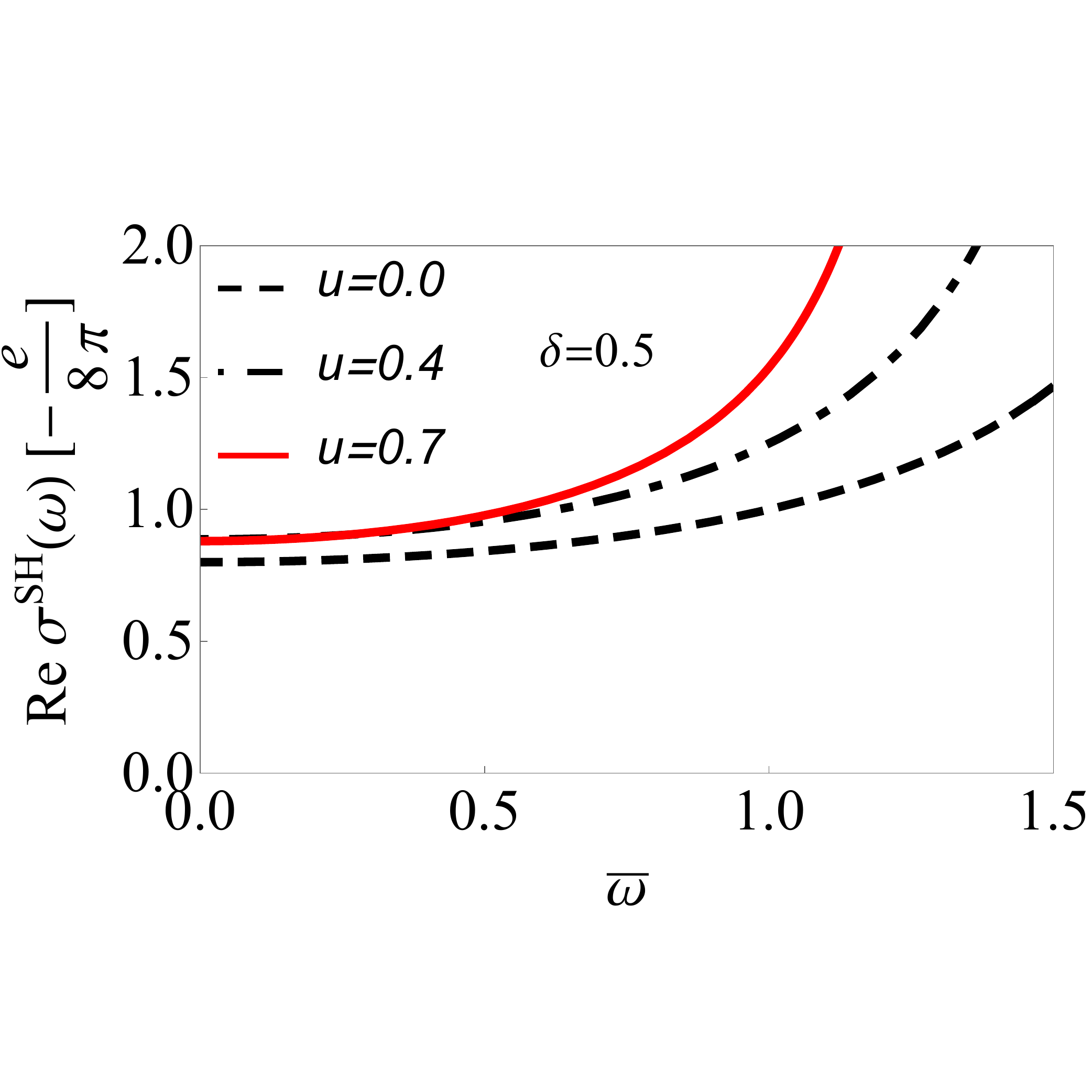}
\hspace{-2em}\includegraphics[width=0.55\linewidth]{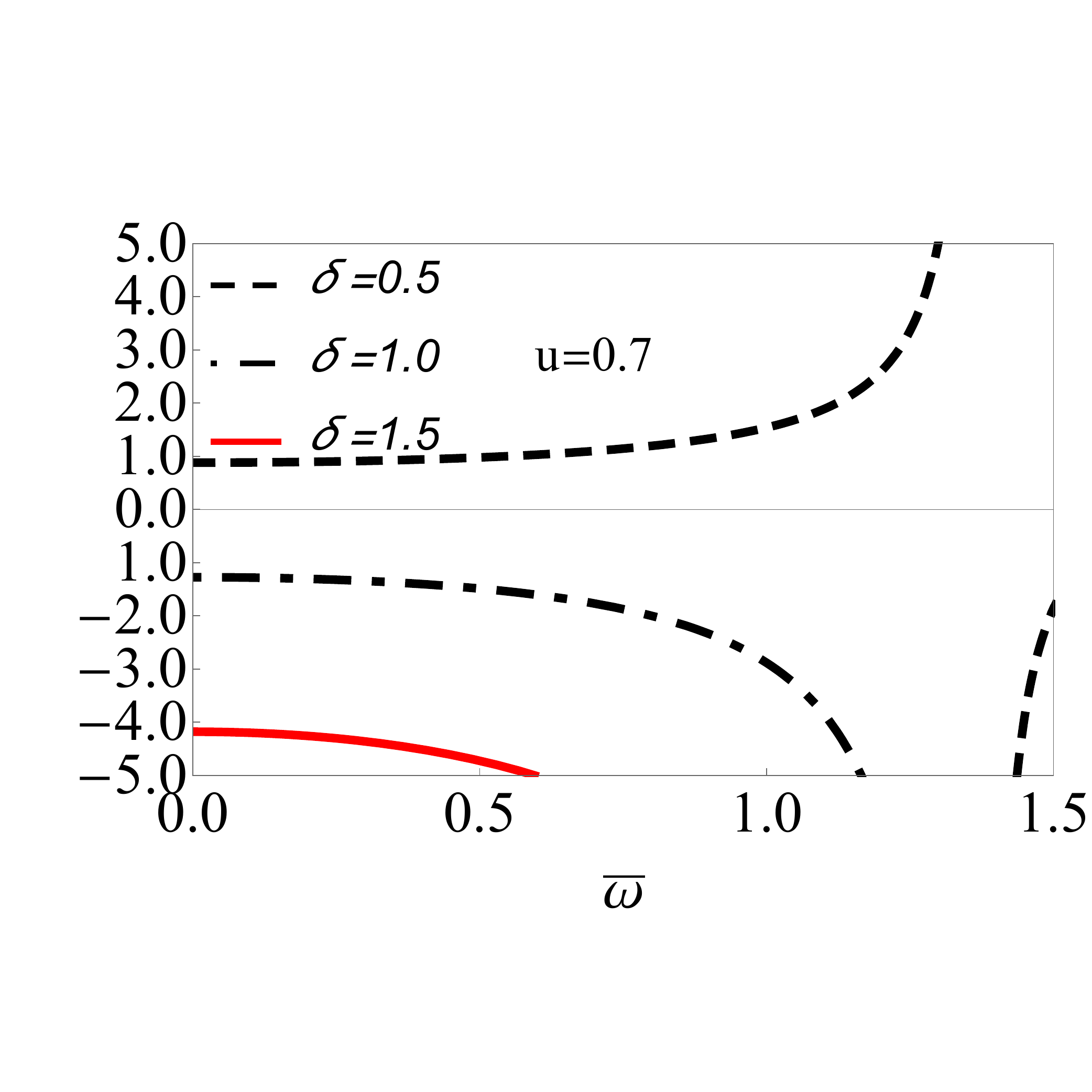}
\end{tabular}
\caption{Interacting dynamical spin Hall conductivity [in units of $-e/(8\pi)$] versus dimensionless frequency $\bar{\omega}=\omega/\sqrt{2m\alpha^2_{\rm R} \epsilon_{\rm F}}$, for $\delta = 0.5$ and different values of the interaction 
		strength $u$ (left), and for $u = 0.7$  and different values of the exchange field
	 $\delta$ (right). We have used $\alpha=0.25$ for the dimensionless spin-orbit coupling strength in both panels. 
\label{fig:Dy-SHC}}
\end{figure}

Similarly, we can obtain the interacting dynamical spin Hall conductivity from Eq.~\eqref{Dy-S-H-C}
\be
\sigma^{\rm SH}_{yx}({\omega})=-\frac{e}{8\pi}\frac{1+\frac{\bar{\omega}^2-4\delta^2}{4\alpha \bar{\omega}}L(\bar{\omega})-\frac{u}{2}\left(\left[1+\frac{\bar{\omega}^2+4\delta^2}{4\alpha \bar{\omega}}L(\bar{\omega})\right]^2-\left[\frac{\delta}{\alpha}L(\bar{\omega})\right]^2\right)}{\left[1-\frac{u}{2}\left(1+\frac{\bar{\omega}^2+4\delta^2}{4\alpha \bar{\omega}}L(\bar{\omega})\right)\right]^2-\left[\frac{u}{2} \frac{\delta}{\alpha}L(\bar{\omega})\right]^2}.
\ee
In Fig.~\ref{fig:Dy-SHC}, the real part of the interacting dynamical spin Hall conductivity is plotted versus frequency for different interaction strength and exchange field values.
%%%%%%%%%%%%Fig.4%%%%%%%%%%%%%%%%%%%

%\end{widetext}

\end{document}